\documentclass[letterpaper, 10 pt, conference]{ieeeconf}%
\usepackage{bm}
\usepackage{amssymb}
\usepackage{epsfig}
\usepackage{subfigure}
\usepackage{amsmath}
\usepackage{graphicx}
\usepackage{epstopdf}
\usepackage{amsfonts}%
\usepackage{indentfirst}
\usepackage{array}
\usepackage[ruled]{algorithm2e}
\usepackage{eufrak}%

\newtheorem{definition}{\rm\textbf{Definition}}
\newtheorem{theorem}{\rm\textbf{Theorem}}

\newtheorem{remark}{\rm\textbf{Remark}}
\providecommand{\U}[1]{\protect\rule{.1in}{.1in}}
\IEEEoverridecommandlockouts
\begin{document}

\title{{\LARGE \textbf{Feasibility-Guided Learning for Robust Control in Constrained Optimal Control Problems}}}
\author{Wei Xiao, Calin A. Belta and Christos G. Cassandras\thanks{This work was supported in
part by the NSF under grants
IIS-1723995, CPS-1446151, ECCS-1931600, DMS-1664644, and CNS-1645681, by ARPA-E’s NEXTCAR program under grant DE-AR0000796, by AFOSR under grant FA9550-19-1-0158, and by the MathWorks. The authors are
with the Division of Systems Engineering and Center for Information and
Systems Engineering, Boston University, Brookline, MA, 02446, USA
\texttt{{\small \{xiaowei, cbelta, cgc\}@bu.edu}}}}
\maketitle

\begin{abstract}

Optimal control problems with constraints ensuring safety and convergence to desired states can be mapped onto a sequence of real time optimization problems through the use of Control Barrier Functions (CBFs) and Control Lyapunov Functions (CLFs). One of the main challenges in these approaches is ensuring the feasibility of the resulting quadratic programs (QPs) if the system is affine in controls. The recently proposed penalty method has the potential to improve the existence of feasible solutions to such problems. In this paper, we further improve the feasibility robustness (i.e.,  feasibility maintenance in the presence of time-varying and unknown unsafe sets) through the definition of a High Order CBF (HOCBF) that works for arbitrary relative degree constraints; this is achieved by a proposed feasibility-guided learning approach. Specifically, we apply machine learning techniques to classify the parameter space of a HOCBF into feasible and infeasible sets, and get a differentiable classifier that is then added to the learning process. The proposed feasibility-guided learning approach is compared with the gradient-descent method on a robot control problem. The simulation results show an improved ability of the feasibility-guided learning approach over the gradient-decent method to determine the optimal parameters in the definition of a HOCBF for the feasibility robustness, as well as show the potential of the CBF method for robot safe navigation in an unknown environment.
\end{abstract}

\thispagestyle{empty} \pagestyle{empty}


\section{Introduction}
\label{sec:intro}

Stabilizing a dynamical system while optimizing a cost function and satisfying safety constraints is a fundamental and challenging problem in control. Typically, such problems include autonomous driving in road traffic and robot safe exploration in unknown environments. When safety becomes critical, it is desired to prioritize the strict satisfaction of constraints instead of optimality. The barrier function method \cite{Aaron2014, Nguyen2016, Lindemann2018, Xiao2019} has been proposed as an approach to this problem.

Barrier functions (BFs) are 
Lyapunov-like functions \cite{Wieland2007}, whose use can be traced back to optimization problems \cite{Boyd2004}. More recently, they have been employed in verification and control, e.g., to prove set invariance \cite{Aubin2009, Prajna2007,Wisniewski2013} and for multi-objective control \cite{Panagou2013}. Control BFs (CBFs) are extensions of BFs for control systems. Recently, it has been shown that CBFs can be combined with
control Lyapunov functions (CLFs) \cite{Sontag1983,Freeman1996,Aaron2012} as constraints to form quadratic programs (QPs) \cite{Galloway2013} for nonlinear control systems that are affine in controls, and these QPs can be solved in real time.  While computationally efficient, the CBF and CLF-based QPs can easily become infeasible in the presence of both stringent safety constraints and tight control limitations, especially for high relative degree systems in a highly dynamical environment. 

The CLF constraints are usually relaxed \cite{Aaron2014} such that they do not conflict with the CBF constraints in the QPs. Recent work showed that rich specifications given in signal temporal logic \cite{Lindemann2018} and linear temporal logic \cite{Nillson2018, Mohit2018} can be translated to constraints and implemented by the CBF method with good solution feasibility if the constraints are with relative degree one. Several approaches to improve feasibility for the CBF and CLF-based QPs on specific applications have been proposed. For the adaptive cruise control (ACC) problem (the system is with relative degree 2) defined in \cite{Aaron2014}, the infeasibility issue is addressed by including the minimum braking distance in the safety constraint. An approximation of the braking distance was used in \cite{Wei2019} for a cooperative optimization control problem with non-linear dynamics. In both cases, an additional complex safety constraint needs to be added. Further, this approach does not scale well for high-dimensional systems.  We recently developed the penalty method \cite{Xiao2019}, which can improve the feasibility of the QPs by penalizing the class $\mathcal{K}$ functions in the definition of a High Order CBF (HOCBF) for an arbitrary relative degree constraint.

The use of machine learning techniques to improve feasibility was recently proposed for legged robots. Feasibility constraints for probabilistic models are learned in \cite{Carpentier2017} based on simplified models. Since the learned constraints are complex, they are simplified by expectation-maximization (EM). Robot footstep limits are modeled as hyper-planes based on success and failure datasets in \cite{Perrin2012}. Reinforcement learning (RL) \cite{Mnih2015} \cite{Mnih2016} has the potential to address the infeasibility issue for optimal control problems, but it is difficult to quantify feasibility as a reward and the optimized parameters may also go to a local infeasible region where a feasible solution could never be found.

 In this paper, we adopt the CBF method to improve the feasibility and feasibility robustness of optimal control problems with stringent safety constraints (usually with high relative degree) and tight control limitations in an unknown environment. The feasibility robustness is defined by the QP feasibility maintenance in the presence of a number of time-varying and unknown unsafe sets. Based on our proposed penalty method from \cite{Xiao2019}, we parameterize a HOCBF, and use the parameters to improve the feasibility of the CBF and CLF-based QPs. Since trajectories of a system may be required to avoid a number of unsafe sets at the same time, we propose the idea of minimizing the value of a HOCBF (usually the distance to an unsafe set) when the corresponding HOCBF constraint first becomes active. In other words, we want the HOCBF constraint to become active as late as possible in the QPs. In this way, the feasibility robustness of the controller with respect to unknown unsafe sets is maximized. The main benefits of maximizing the robustness lie in the fact that the QP feasiblity can be maintained when the unsafe sets are unknown and with detection noise, as will be shown later. Another contribution of this paper is to put forward a feasibility-guided method to learn the optimal parameters in a HOCBF corresponding to a specific type of unsafe set such that the robustness is maximized. We compare the proposed feasibility-guided method with the gradient-descent method with results showing improved controller robustness in a robot control problem.

This paper is structured as follows. In Sec. \ref{sec:pre}, we give preliminaries on HOCBFs and CLFs. In Sec. \ref{sec:prob}, we formulate an optimal control problem with safety constraints and control limitations. The framework of learning the optimal penalties and powers for a specific type of unsafe set is given in Sec. \ref{sec:learn}. We provide simulations and comparisons in Sec. \ref{sec:sim},  and conclude with final remarks and directions for future work in Sec. \ref{sec:con}.

\section{Preliminaries}

\label{sec:pre}
\begin{definition} \label{def:classk}
	({\it Class $\mathcal{K}$ function} \cite{Khalil2002}) A continuous function $\alpha:[0,a)\rightarrow[0,\infty), a > 0$ is said to belong to class $\mathcal{K}$ if it is strictly increasing and $\alpha(0)=0$. 
\end{definition}

Consider an affine control system of the form
\begin{equation} \label{eqn:affine}
\dot {\bm{x}} = f(\bm x) + g(\bm x)\bm u
\end{equation}
where  $\bm x\in \mathbb{R}^n$, $f:\mathbb{R}^n\rightarrow \mathbb{R}^n$ and $g:\mathbb{R}^n \rightarrow \mathbb{R}^{n\times q}$ are globally Lipschitz, and $\bm u\in U \subset \mathbb{R}^q$ ($U$ denotes the control constraint set). Solutions $\bm x(t)$ of (\ref{eqn:affine}), starting at $\bm x(0)$ (we set the initial time to 0 without loss of generality), $t\geq 0$, are forward complete.

\begin{definition} \label{def:forwardinv}
	A set $C\subset\mathbb{R}^n$ is forward invariant for system (\ref{eqn:affine}) if its solutions starting at any $\bm x(0) \in C$ satisfy $\bm x(t)\in C$ for $\forall t\geq 0$.
\end{definition} 

\begin{definition} \label{def:relative}
	({\it Relative degree})
	The relative degree of a continuously differentiable function $b:\mathbb{R}^n\rightarrow \mathbb{R}$ with respect to system (\ref{eqn:affine}) is the number of times we need to differentiate it along its dynamics until the control $\bm u$ explicitly shows in the corresponding derivative. 
\end{definition}

In this paper, since function $b$ is used to define a constraint $b(\bm x)\geq 0$, we will also refer to the relative degree of $b$ as the relative degree of the constraint.

For a constraint $b(\bm x)\geq 0$ with relative degree $m$, $b: \mathbb{R}^n \rightarrow \mathbb{R}$, and $\psi_0(\bm x) := b(\bm x)$, we define a sequence of functions  $\psi_1: \mathbb{R}^n \rightarrow \mathbb{R}, \psi_2: \mathbb{R}^n  \rightarrow \mathbb{R},\dots, \psi_{m}: \mathbb{R}^n  \rightarrow \mathbb{R}$:
\begin{equation} \label{eqn:functions}
\begin{aligned}
\psi_i(\bm x) := \dot \psi_{i-1}(\bm x) + \alpha_i(\psi_{i-1}(\bm x)),i\in\{1,2,\dots,m\},
\end{aligned}
\end{equation}
where $\alpha_i(\cdot),i\in\{1,2,\dots,m\}$ denotes a differentiable class $\mathcal{K}$ function.

We further define a sequence of sets $C_1, C_2,\dots, C_{m}$ associated with (\ref{eqn:functions}) in the form:
\begin{equation} \label{eqn:sets}
\begin{aligned}
C_i := \{\bm x \in \mathbb{R}^n: \psi_{i-1}(\bm x) \geq 0\}, i\in\{1,2,\dots,m\}.
\end{aligned}
\end{equation}

\vspace{2ex}
\begin{definition} \label{def:hocbf}
	({\it High Order Control Barrier Function (HOCBF)} \cite{Xiao2019}) Let $C_1, C_2,\dots, C_{m}$ be defined by (\ref{eqn:sets}) and $\psi_1(\bm x), \psi_2(\bm x),\dots, \psi_{m}(\bm x)$ be defined by (\ref{eqn:functions}). A function $b: \mathbb{R}^n\rightarrow \mathbb{R}$ is a high order control barrier function (HOCBF) of relative degree $m$ for system (\ref{eqn:affine}) if there exist differentiable class $\mathcal{K}$ functions $\alpha_1,\alpha_2,\dots, \alpha_{m}$ such that
	\begin{equation}\label{eqn:constraint}
	\begin{aligned}
	L_f^{m}b(\bm x) + L_gL_f^{m-1}b(\bm x)\bm u + O(b(\bm x)) + \alpha_m(\psi_{m-1}(\bm x)) \geq 0,
	\end{aligned}
	\end{equation}
	for all $\bm x\in C_1 \cap C_2\cap,\dots, \cap C_{m}$. In (\ref{eqn:constraint}), $L_f, L_g$ denote Lie derivatives along $f$ and $g$, respectively, $O(\cdot)$ denotes the remaining Lie derivatives along $f$ with degree less than or equal to $m-1$ (omitted for simplicity, see \cite{Xiao2019}).
\end{definition}

\vspace{2ex}
Given a HOCBF $b$, we define the set of all control values that satisfy (\ref{eqn:constraint}) as:
\begin{equation}
\begin{aligned}
K_{cbf} = \{\bm u\in U: L_f^{m}b(\bm x) + L_gL_f^{m-1}b(\bm x)\bm u \\ + O(b(\bm x))  + \alpha_m(\psi_{m-1}(\bm x)) \geq 0\}
\end{aligned}
\end{equation}

\vspace{2ex}
\begin{theorem} \label{thm:hocbf}
	(\cite{Xiao2019}) Given a HOCBF $b(\bm x)$ from Def. \ref{def:hocbf} with the associated sets $C_1, C_2,\dots, C_{m}$ defined by (\ref{eqn:sets}), if $\bm x(0) \in C_1 \cap C_2\cap,\dots,\cap C_{m}$, then any Lipschitz continuous controller $\bm u(t)\in K_{cbf}, \forall t\geq 0$ renders
	$C_1$ $\cap  C_2\cap,\dots, \cap C_{m}$ forward invariant for system (\ref{eqn:affine}).
\end{theorem}

\begin{definition}  \label{def:clf}
	({\it Control Lyapunov function (CLF)} \cite{Aaron2012}) A continuously differentiable function $V: \mathbb{R}^n\rightarrow \mathbb{R}$ is a globally and exponentially stabilizing control Lyapunov function (CLF) for system (\ref{eqn:affine}) if there exist constants $c_1 >0, c_2>0, c_3>0$ such that
	\begin{equation}
	c_1||\bm x||^2 \leq V(\bm x) \leq c_2 ||\bm x||^2
	\end{equation}
	\begin{equation}\label{eqn:clfp}
	\underset{u\in U}{inf} \lbrack L_fV(\bm x)+L_gV(\bm x) \bm u + c_3V(\bm x)\rbrack \leq 0.
	\end{equation}
	for $\forall \bm x\in \mathbb{R}^n$.
\end{definition}

\begin{theorem} \label{thm:clf}
	(\cite{Aaron2012}) Given an exponentially stabilizing CLF $V$ as in Def. \ref{def:clf}, any Lipschitz continuous controller $ \bm u \in K_{clf}(\bm x)$, with
	$$K_{clf}(\bm x) := \{\bm u\in U: L_fV(\bm x)+L_gV(\bm x) \bm u + c_3V(\bm x) \leq 0\},$$
	exponentially stabilizes system (\ref{eqn:affine}) to its zero dynamics (defined by the dynamics of the internal part if we transform the system to standard form and set the output to zero \cite{Khalil2002}). Note that (\ref{eqn:clfp}) can be relaxed by adding a relaxation at its right-hand side \cite{Aaron2012}.
\end{theorem}

Recent works \cite{Aaron2014}, \cite{Lindemann2018}, \cite{Nguyen2016} combine CBFs and CLFs with quadratic costs to form optimization problems. Time is discretized and an optimization problem with constraints given by CBFs and CLFs is solved at each time step. Note that these constraints are linear in control since the state is 
fixed at the value at the beginning of the interval, and therefore the optimization problem is a quadratic program (QP). The optimal control obtained by solving the QP is applied 
at the current time step and held constant for the whole interval. The dynamics (\ref{eqn:affine}) are updated, and the procedure is repeated. This method works conditioned on the fact that the QP is always feasible. We will show how we can further improve the QP feasibility by maximizing the feasibility robustness (will be formally defined in the next section) of the controller with respect to unknown unsafe sets in this paper.

\section{Problem Formulation}
\label{sec:prob}

Consider an optimal control problem for system (\ref{eqn:affine}) with the cost defined as:
\begin{equation}\label{eqn:gcost}
\min_{\bm u(t)}J(\bm u(t)) = \int_{0}^{t_f}\mathcal{C}(||\bm u(t)||)dt,
\end{equation} 
where $||\cdot||$ denotes the 2-norm of a vector; $t_f$ denotes the final time; and $\mathcal{C}(\cdot)$ is a strictly increasing function of its argument. 

$\textbf{State convergence}$: We want the state of system (\ref{eqn:affine}) to converge to a point $\bm K\in\mathbb{R}^n$, i.e.,
\begin{equation} \label{eqn:target}
||\bm x(t) - \bm K|| \leq \xi, \forall t\in[t',t_f],
\end{equation}
where $\xi > 0$ is arbitrarily small and $t'\in[0,t_f]$.

$\textbf{Constraint 1}$ (Unsafe Sets): Let $S_o$ denote a set of unsafe sets.
System (\ref{eqn:affine}) should always avoid all unsafe regions (obstacles) $j\in S_o$, i.e.,
\begin{equation} \label{eqn:obstacle}
h_j(\bm x(t))\geq 0, \forall t\in[0,t_f].
\end{equation}
where $h_j: \mathbb{R}^n\rightarrow\mathbb{R}, \forall j\in S_o$ is continuously differentiable.

A HOCBF constraint for (\ref{eqn:obstacle}) becomes active when a control $\bm u$ makes inquality (\ref{eqn:constraint}) become an equality for $b = h_j$.

$\textbf{Feasibility robustness}$: The feasibility robustness of a controller with respect to a constraint (\ref{eqn:obstacle}) can be quantified by the value of $h_j(\bm x(t_a))$ (the value of $h_j(\cdot)$ usually denotes the distance to the unsafe set $j\in S_o$) when the HOCBF constraint (\ref{eqn:constraint}) for (\ref{eqn:obstacle}) first becomes active (and active afterwards) at $t_a\in[0,t_f]$. In order to maximize the feasibility robustness, we need to minimize
\begin{equation} \label{eqn:free}
\min_{t_a} h_j(\bm x(t_a)), j\in S_o.
\end{equation}

As an example, consider the adaptive cruise control problem \cite{Xiao2019}. The distance $z(t)$ between the controlled vehicle and the vehicle in front (both vehicles have double integrator dynamics and control constraints) should be greater than a constant $\delta > 0$, i.e., $z(t) \geq \delta, \forall t\geq 0$. Then we can define a HOCBF $h(\bm x) := z(t) - \delta$ ($m = 2$ in Def. \ref{def:hocbf} since the relative degree of $h(\cdot)$ is 2 for double integrator dynamics) for this safety constraint, and any control input should satisfy the HOCBF constraint (\ref{eqn:constraint}). If the HOCBF constraint is first active at $t_a$ and the value of $h(\bm x(t_a))$ is very small (while the control constraints should be always satisfied), i.e., the distance between these two vehicles is small, then the controller (from the QPs) is less constrained (before the HOCBF constraint becomes active) and robust to perturbations (such as noise). Thus, the feasibility robustness of the controller is improved and we wish to solve $\min_{t_a} h(\bm x(t_a))$.

\begin{remark}
	There are three main advantages of maximizing the feasibility robustness of the controller. (i) The QPs are more likely to become feasible since fewer (HOCBF) constraints will become active when a system gets close to a number of unsafe sets; (ii) In an {\it unknown} environment, the controller obtained through the QPs is more robust to the change of environment and the detection of unknown unsafe sets since the corresponding HOCBF constraints only work (become active) when a system gets close to these unsafe sets. If the corresponding HOCBF constraints become active before the unsafe sets are detected, the system will fail to avoid these unsafe sets (i.e., QPs become infeasible). (iii) There is higher probability to find a better solution (e.g., energy optimal) if the feasibility robustness is maximized since the QP solutions are less constrained.
\end{remark}

$\textbf{Constraint 2}$ (State Limitations): Assume we have a set of constraints on the state of system (\ref{eqn:affine}) in the form:
\begin{equation}\label{eqn:state}
\bm x_{min} \leq \bm x(t)\leq \bm x_{max}, \forall t\in[0,t_f],
\end{equation}
where $\bm x_{max}: = (x_{max,1},x_{max,2},\dots,x_{max,n})\in \mathbb{R}^n$ and $\bm x_{min}: = (x_{min,1},x_{min,2},\dots,x_{min,n})\in \mathbb{R}^n$ denote the maximum and minimum state vectors, respectively, and the inequality is interpreted componentwise.

$\textbf{Constraint 3}$ (Control limitations): Assume we have a set of constraints on control inputs of system (\ref{eqn:affine}) in the form:
\begin{equation}\label{eqn:control}
\bm u_{min}\leq \bm u(t)\leq \bm u_{max}, \forall t\in[0,t_f].
\end{equation}
where $\bm u_{min}\in\mathbb{R}^q$ and $\bm u_{max}\in\mathbb{R}^q$ denote the minimum and maximum
control input vectors, respectively (i.e., the constraint set $U$ in (\ref{eqn:affine}) is rectangular).

A control policy for system (\ref{eqn:affine}) is $\bm {feasible}$ if constraints (\ref{eqn:obstacle}), (\ref{eqn:state}) and (\ref{eqn:control}) are satisfied. In this paper, we consider the following problem:

\textbf{Problem:}
	Find a feasible control policy for system (\ref{eqn:affine}) such that cost (\ref{eqn:gcost}) is minimized, robustness is maximized (i.e., (\ref{eqn:free}) is minimized),
    constraints 1, 2, 3 ((\ref{eqn:obstacle}), (\ref{eqn:state}) and (\ref{eqn:control}))
	are strictly satisfied,
	and state convergence (\ref{eqn:target}) is satisfied with the smallest possible $\xi$ and $t'$.

\textbf{Approach:} The robustness objective (\ref{eqn:free}) depends on the time $t_a$, while $t_a$ is determined once a HOCBF in the above problem is defined. Therefore, we need to consider objective (\ref{eqn:free}) in the definition of a HOCBF. We break the above problem into two sub-problems: (i) objective (\ref{eqn:gcost}) subject to (\ref{eqn:obstacle}), (\ref{eqn:state}), (\ref{eqn:control}) and (\ref{eqn:target}) that is solved with the QP-based method introduced at the end of Sec. \ref{sec:pre}; (ii) objective (\ref{eqn:free}) after solving sub-problem (i) $\forall t\in[0,t_f]$.

\section{Learning to Increase Feasibility Robustness}
\label{sec:learn}

In this section, we introduce how to learn the optimal parameters in the definition of a HOCBF such that the feasibility robustness of the controller with respect to unknown unsafe sets is maximized, i.e., how to reformulate sub-problem (i), (ii) introduced at the end of the last section. We define unsafe sets as being of the same ``type'' if they have the same geometry. For example, circular unsafe sets are the same type if they have the same radius but different locations. Let $S_t\subseteq S_o$ denote the index set of all the unsafe set types in $S_o$.

\subsection{HOCBF and CLF-based QP (sub-problem (i))}
\label{sec:subp1}

The approach to sub-problem (i) is based on 
partitioning the time interval $[0,t_f]$  into a set of equal time intervals $\{[0, \Delta t), [\Delta t,2\Delta t),\dots\}$, where $\Delta t > 0$. In each interval $[\omega \Delta t, (\omega+1) \Delta t)$ ($\omega = 0,1,2,\dots$), we assume the control is constant (i.e., the overall control will be piece-wise constant).
Then at $t = \omega \Delta t$, we solve 
\begin{equation} \label{eqn:obj}
\min_{\bm u(t),\delta(t)} \mathcal{C}(||\bm u(t)||) + p\delta^2(t)
\end{equation}
subject to (\ref{eqn:control}), the CLF constraint (\ref{eqn:clfp}) for (\ref{eqn:target}) (by defining a CLF for (\ref{eqn:target}) such that a CLF constraint similar to (\ref{eqn:clfp} is satisified) and the HOCBF constraints (\ref{eqn:constraint}) corresponding to (\ref{eqn:obstacle}) and (\ref{eqn:state}), where $p > 0$ is a penalty on the relaxation $\delta(t)$ ($\delta(t)$ is a relaxation variable that replaces 0 on the right-hand side of (\ref{eqn:clfp})). Since the state is kept constant at its value at the beginning of the interval, the above optimization problem is a QP, which can easily become infeasible. In the rest of the paper, we show how we can use machine learning techniques in finding the optimal parameters in the definition of a HOCBF such that the feasibility robustness is maximized. 

\subsection{The Penalty Method}

To improve the feasibility \cite{Xiao2019} of the problem (\ref{eqn:obj}), we add penalties on the class $\mathcal{K}$ functions $\alpha_1(\cdot),\alpha_2(\cdot),\dots, \alpha_{m}(\cdot)$ in (\ref{eqn:functions}) in the definition of a HOCBF $b(\bm x)$. 
In the set of class $\mathcal{K}$ functions that consist of power functions, we explictly rewrite (\ref{eqn:functions}) as
\begin{equation} \label{eqn:pqfunctions}
\begin{aligned}
\psi_0(\bm x) :=& b(\bm x)\\
\psi_1(\bm x) :=& \dot \psi_0(\bm x) + p_1\psi_0^{q_1}(\bm x),\\
&\vdots\\
\psi_m(\bm x) :=& \dot \psi_{m-1}(\bm x) + p_m\psi_{m-1}^{q_m}(\bm x),
\end{aligned}
\end{equation}
where $p_1 >0, p_2 > 0,\dots,p_m > 0$ and $q_1 >0, q_2 > 0,\dots,q_m > 0$.

For each type of unsafe set $j\in S_t$, we consider an arbitrary location for it and get an unsafe set constraint $h_j(\bm x(t))\geq 0$ (similar to (\ref{eqn:obstacle})). Let $\bm p := (p_1, p_2,\dots,p_m)$, $\bm q := (q_1, q_2,\dots,q_m)$. We know from \cite{Xiao2019} that the values of $q_1, q_2,\dots,q_m$ affect the feasibility region of (\ref{eqn:obj}), as well as what time the HOCBF constraint (\ref{eqn:constraint}) will be active, i.e., we can rewrite $h_j(\bm x(t_a))$ as $h_j(\bm x(t_a),\bm p, \bm q)$. Let $\mathcal{D}_j(\bm p, \bm q):= h_j(\bm x(t_a),\bm p, \bm q)$ (since $h_j(\bm x(t_a),\bm p, \bm q)$ is fixed once $\bm p, \bm q$ are given, $h_j(\cdot)$ does not actually depend on $\bm x(t_a)$). We reformulate (\ref{eqn:free}) as
\begin{equation}\label{eqn:racost}
\min_{\bm p,\bm q} \mathcal{D}_j(\bm p, \bm q), j\in S_t.
\end{equation}
We can view the minimization of ${D}_j(\bm p, \bm q)$ as the maximization of the feasibility robustness that depends on $\bm p, \bm q$.

Then, we need to find the optimal $\bm p$ and $\bm q$ that minimize (\ref{eqn:racost}) for each unsafe set type $j\in S_t$. However, this optimization problem is hard to solve. We will introduce an approach using machine learning techniques in the following section.

\subsection{Feasibility-Guided Optimization (sub-problem (ii))}

The optimization problem (\ref{eqn:racost}) is a typical problem that can be solved with reinforcement learning approaches. However, most of the $\bm p, \bm q$ values result in infeasible solutions of problem (\ref{eqn:obj}), which makes (\ref{eqn:racost}) difficult to solve. Therefore, we need to first solve the infeasiblity problem of sub-problem (i).
We randomly sample $\bm p, \bm q$ values over their domain, and for each set of $\bm p, \bm q$ values, we solve problem (\ref{eqn:obj}) until the state convergence (\ref{eqn:target}) is achieved. If problem (\ref{eqn:obj}) (the QPs) is feasible at all times, then we label this set of $\bm p, \bm q$ values (as a whole) as $+1$, otherwise, we label it as $-1$. Eventually, we get sets of feasible and infeasible $\bm p, \bm q$ points. Then we can apply a machine learning technique (such as support vector machine, deep neural network etc.) to classify these two sets and get a continuously differentiable hypersurface:
\begin{equation}\label{eqn:hyper}
\mathfrak{H}_j:\mathbb{R}^{2m}\rightarrow \mathbb{R},
\end{equation}
where 
\begin{equation}\label{eqn:hyperconst}
\mathfrak{H}_j(\bm p,\bm q)\geq 0
\end{equation}
denotes the set of $\bm p,\bm q$ values (as a whole) which leads to the feasible solution of QPs (\ref{eqn:obj}), i.e., the feasibility constraint for the set of $\bm p, \bm q$ values associated with the QPs (\ref{eqn:obj}). We can use a HOCBF to enforce (\ref{eqn:hyperconst}) if $\bm p,\bm q$ are state variables of a system, which motivates us to define dynamics for $\bm p,\bm q$, as shown later.

Based on the feasiblity classification hypersurface (\ref{eqn:hyper}), we look further to optimize (\ref{eqn:racost}), i.e., we consider (\ref{eqn:racost}) subject to (\ref{eqn:hyperconst}). However, the learned hypersurface (\ref{eqn:hyper}) is generally complex, and thus makes this optimization problem very hard to solve. We use the following approach to simplify this optimization problem.

We start at some feasible $\bm p_0\in\mathbb{R}^m,\bm q_0\in\mathbb{R}^m$ to search for the optimal $\bm p,\bm q$ values. Since the determination of the optimal $\bm p, \bm q$ is a dynamic process, we define the gradient (dynamics) for $\bm p,\bm q$ (as the variations of $\bm p,\bm q$ that are controlled), i.e., we have
\begin{equation}\label{eqn:assist}
( \dot {\bm p}(\mathfrak{t}), \dot  {\bm q}(\mathfrak{t})) = \bm \nu(\mathfrak{t}),\dot {\bm p}(\mathfrak{t}_0) = \bm p_0, \dot  {\bm q}(\mathfrak{t}_0) = \bm q_0,
\end{equation}
where $\bm \nu \in \mathbb{R}^{2m}$ denotes a controllable input vector in the dynamic process constructed in order to determine the optimal $\bm p, \bm q$. $\mathfrak{t}$ denotes the dynamic process time for the optimization of (\ref{eqn:racost}), which is different and independent from $t$ in the system (\ref{eqn:affine}) and problem (\ref{eqn:obj}). $\mathfrak{t}_0\in\mathbb{R}$ denotes the initial time.

Considering feasibility of the problem (\ref{eqn:obj}), the dynamic process (determined by $\bm \nu$) should be subjected to (\ref{eqn:assist}) and (\ref{eqn:hyperconst}). Since we want to find the control $\bm\nu$ such that the resulting $\bm p, \bm q$ (determined by $\bm\nu$) always lead to the feasible solution of QPs (\ref{eqn:obj}) with the CBF method, i.e., we need to take the derivative of (\ref{eqn:hyper}), we minimize the derivative of (\ref{eqn:racost}) (the fastest decreasing direction of the value of $\mathcal{D}_j(\bm p, \bm q)$ in (\ref{eqn:racost})) in the dynamic process to make $\bm\nu$ also show up in the cost function. As long as the derivative of (\ref{eqn:racost}) is negative, we make sure that (\ref{eqn:racost}) is decreasing in each time step (by discretizing $\mathfrak{t}$ similar to sub-problem (i)).

By taking the derivative of (\ref{eqn:racost}) with respect to $\mathfrak{t}$, we have
\begin{equation}\label{eqn:deracost}
\begin{aligned}
\frac{ d\mathcal{D}_j(\bm p(\mathfrak{t}), \bm q(\mathfrak{t}))}{d\mathfrak{t}}& = \frac{ d\mathcal{D}_j( \bm p(\mathfrak{t}), \bm q(\mathfrak{t}))}{d(\bm p(\mathfrak{t}), \bm q(\mathfrak{t}))}\frac{d(\bm p(\mathfrak{t}), \bm q(\mathfrak{t}))}{d\mathfrak{t}}
\\&=\frac{ d\mathcal{D}_j(\bm p(\mathfrak{t}), \bm q(\mathfrak{t}))}{d(\bm p(\mathfrak{t}), \bm q(\mathfrak{t}))} \bm \nu.
\end{aligned}
\end{equation}

The relative degree of the feasibility constraint (\ref{eqn:hyperconst}) with respect to (\ref{eqn:assist}) is 1. We then use a  HOCBF with $m = 1$ (as in Def. \ref{def:hocbf}) to enforce (\ref{eqn:hyperconst}) and find a control $\bm\nu$ that can satisfy (\ref{eqn:hyperconst}) in the dynamic process:
\begin{equation}\label{eqn:hyperder}
\frac{d\mathfrak{H}_j(\bm p,\bm q)}{d(\bm p, \bm q)}\bm \nu + \alpha_1(\mathfrak{H}_j(\bm p,\bm q)) \geq 0,
\end{equation}
where $\alpha_1(\cdot)$ is a class $\mathcal{K}$ function as in Def. \ref{def:hocbf} (the definition of $\psi_1(\cdot)$ in (\ref{eqn:functions})).
Any control input $\bm\nu$ that satisfies (\ref{eqn:hyperder}) implies that the resulting $\bm p, \bm q$ (determined by $\bm\nu$) lead to a feasible solution of QPs (\ref{eqn:obj}) in the dynamic process.

Then, we reformulate sub-problem (ii) by the dynamic process (\textbf{feasibility-guided optimization} (FGO)).  We use the approach introduced as in Sec. \ref{sec:subp1} to solve the dynamic process, i.e., we discretize $\mathfrak{t}$, at each $\mathfrak{t} = \omega \Delta \mathfrak{t}, \omega\in\{0,1,\dots\}$ ($\Delta \mathfrak{t}>0$ denotes the discretization constant), and we solve
\begin{equation}\label{eqn:dejracost}
\min_{\bm \nu(\mathfrak{t})} \frac{ d\mathcal{D}_j(\bm p(\mathfrak{t}), \bm q(\mathfrak{t}))}{d(\bm p(\mathfrak{t}), \bm q(\mathfrak{t}))} \bm \nu(\mathfrak{t})
\end{equation}
subject to (\ref{eqn:hyperder}), (\ref{eqn:assist}). Then update (\ref{eqn:assist}) for $\mathfrak{t}\in(\omega \Delta \mathfrak{t}, (\omega+1) \Delta \mathfrak{t})$ with $\bm\nu^*(\mathfrak{t})$. Note that in the last equation, $\frac{ d\mathcal{D}_j(\bm p(\mathfrak{t}), \bm q(\mathfrak{t}))}{d(\bm p(\mathfrak{t}), \bm q(\mathfrak{t}))}$ is a vector of dimension $1\times 2m$, while $\bm \nu$ is a vector of dimension $2m\times 1$. Therefore, the cost function in the last equation is a scalar function of $\bm \nu$.

   The optimization problem (\ref{eqn:dejracost}) is a linear program (LP) (to determine $\bm\nu$) at each time step for each initial $\bm p, \bm q$ (we need to reset $\mathfrak{t}$ for each set of initial $\bm p, \bm q$ values). Without any constraint on $\bm\nu$, the LP (\ref{eqn:dejracost}) is ill-posed because it leads to unbounded solutions. In fact, the value of $\bm \nu$ determines the search step length of the dynamic process, and we want to limit this step length. Otherwise, the solution of the LP at each step is infinity (i.e., the dynamic process search step length is infinity, and fails to work). Therefore, we add limitations to $\bm\nu$ for the LP (\ref{eqn:dejracost}):
\begin{equation}\label{eqn:assistcontrol}
\bm \nu_{min} \leq \bm \nu \leq \bm \nu_{max}.
\end{equation}
where $\bm\nu_{min} < \bm 0, \bm\nu_{max}> \bm 0$ (interpreted componentwise), $\bm 0 \in\mathbb{R}^{2m}$.

After adding (\ref{eqn:assistcontrol}) to (\ref{eqn:dejracost}), the dynamic process search step length will become bounded. Although there are control limitations on $\bm\nu$, the resulting LP from the optimization (\ref{eqn:dejracost}) is always feasible since the relative degree of (\ref{eqn:hyperconst}) with respect to (\ref{eqn:assist}) is 1.

Note that in (\ref{eqn:dejracost}), we have 
$$\frac{ d\mathcal{D}_j}{d(\bm p, \bm q)} = (\frac{\partial\mathcal{D}_j}{\partial p_1},\dots, \frac{\partial\mathcal{D}_j}{\partial p_m}, \frac{\partial\mathcal{D}_j}{\partial q_1},\dots, \frac{\partial\mathcal{D}_j}{\partial q_m})$$
and we also need to evaluate $\frac{\partial\mathcal{D}_j}{\partial p_1} ,\dots, \frac{\partial\mathcal{D}_j}{\partial p_m}, $$ \frac{\partial\mathcal{D}_j}{\partial q_1},\dots, \frac{\partial\mathcal{D}_j}{\partial q_m}$ at each time step (i.e., evaluate the coefficients of the cost function (\ref{eqn:dejracost})). 

We present the FGO algorithm in Alg. \ref{alg:fgo}. For each step of the FGO algorithm from Alg. \ref{alg:fgo}, the following four conditions may terminate the algorithm: $(i)$ the problem (\ref{eqn:obj}) becomes infeasible (since the hypersurface (\ref{eqn:hyper}) from the machine learning techniques cannot ensure 100\% classification accuracy), $(ii)$ the evaluated values of $\frac{\partial\mathcal{D}_j}{\partial p_1} ,\dots, \frac{\partial\mathcal{D}_j}{\partial p_m}, $$ \frac{\partial\mathcal{D}_j}{\partial q_1},\dots, \frac{\partial\mathcal{D}_j}{\partial q_m}$ are all 0, $(iii)$ the objective function value of (\ref{eqn:racost}) is greater than the current known minimum value. $(iv)$ the iteration time exceeds some $N\in\mathbb{N}$. 

If we consider Alg. \ref{alg:fgo} without the constraint (\ref{eqn:hyperder}), then we have the commonly used gradient descent (GD) algorithm. The FGO algorithm is more conservative compared with GD since the solution searching path is guided by the feasibility of (\ref{eqn:obj}). We can apply GD one step forward whenever the FGO algorithm terminates to alleviate this limitation, which is shown in the last part of Alg. \ref{alg:fgo}.

\begin{algorithm}
	\caption{FGO algorithm} \label{alg:fgo}
	\KwIn{Constraints (\ref{eqn:obstacle}) (\ref{eqn:target}), system (\ref{eqn:affine}) with (\ref{eqn:control}), $N$}
	\KwOut{$\bm p^*,\bm q^*, \mathcal{D}_{min}$}
	
	Sample $\bm p,\bm q$ in the definition of the HOCBF\;
	Discard samples that do not meet the initial conditions of HOCBF constraint (\ref{eqn:constraint})\;
	Solve (\ref{eqn:obj}) for each sample for $t\in [0,t_f]$ and label all samples\;
	Use machine learning to find classifier (\ref{eqn:hyper})\;
	Pick a feasible $\bm p_0, \bm q_0$, $\mathcal{D}_{min}:=\mathcal{D}_j(\bm p_0, \bm q_0)$, iter. = 1\;
	\While{iter.++ $\leq N$}{
		Evaluate $\frac{\partial\mathcal{D}_j}{\partial p_1} ,\dots, \frac{\partial\mathcal{D}_j}{\partial p_m}, $$ \frac{\partial\mathcal{D}_j}{\partial q_1},\dots, \frac{\partial\mathcal{D}_j}{\partial q_m}$ at $\bm p_0, \bm q_0$\;
		\eIf{$\frac{\partial\mathcal{D}_j}{\partial p_k},\frac{\partial\mathcal{D}_j}{\partial q_k}, \forall k\in\{1,2,\dots,m\}$ is infeasible}
		{Jump to the very begining of the loop\;}
		{
			$\frac{\partial\mathcal{D}_j}{\partial p_k} = 0,\frac{\partial\mathcal{D}_j}{\partial q_k} = 0$ if $\exists k\in\{1,2,\dots,m\}$ such that $\frac{\partial\mathcal{D}_j}{\partial p_k},\frac{\partial\mathcal{D}_j}{\partial q_k}$ is infeasible to evaluate\;
		}
		Solve optimization (\ref{eqn:dejracost}) and get new $\bm p, \bm q$\;
		Solve problem (\ref{eqn:obj}) with $\bm p, \bm q$\;
		\eIf{(\ref{eqn:obj}) is feasible}
		{   
			\eIf{$\mathcal{D}_{min}\geq\mathcal{D}_j(\bm p, \bm q)$}
			{
			$\mathcal{D}_{min} = \mathcal{D}_j(\bm p, \bm q)$, $\bm p_0 = \bm p, \bm q_0 = \bm q$\;
			}
		    {break\;}
		}
		{Solve optimization (\ref{eqn:dejracost}) without (\ref{eqn:hyperder}) and get new $\bm p, \bm q$\;
		Solve problem (\ref{eqn:obj}) with $\bm p, \bm q$\;
		\eIf{$\mathcal{D}_{min}\geq\mathcal{D}_j(\bm p, \bm q)$}
		{
			$\mathcal{D}_{min} = \mathcal{D}_j(\bm p, \bm q)$, $\bm p_0 = \bm p, \bm q_0 = \bm q$\;
		}
	    {break\;}
	    }

	}
	$\bm p^* = \bm p_0, \bm q^* = \bm q_0$\;
\end{algorithm}

\subsection{Feasibility Generalization}

The feasibility and feasibility robustness of the controller for problem (\ref{eqn:obj}) is sensitive to the ``shape'' of the unsafe sets, but not to the ``location''. For example, the location of a circular obstacle does not affect the feasibility and feasibility robustness of the controller for a robot, but the geometry of this circular obstacle does. In this case, we do not need to know the exact location of the obstacle. If we have learned feasibility for a specific location obstacle and get optimal $\bm p^*, \bm q^*$ with the FGO algorithm, then the optimal $\bm p^*, \bm q^*$ apply to other located obstacles of the same geometry.

In the case that we know the type of unsafe sets but not the locations, we can learn feasibility and robustness for each type of unsafe set given an arbitrary location with the FGO algorithm. Since the initial system condition may also affect the feasibility of problem (\ref{eqn:obj}), we may learn the optimal  $\bm p^*, \bm q^*$ under the worst initial conditions (e.g., with maximum obstacle-approaching speed for a robot), and then these optimal  $\bm p^*, \bm q^*$ may also apply to other initial conditions. For example, we may set the initial heading angle (as well as the target heading angle) of a robot so as to initially pass through the center of the circle obstacle and set the speed to its maximum speed. Once the optimal  $\bm p^*, \bm q^*$ are found under this conditon, they may also be applied to other conditions.

In an unknown environment, system (\ref{eqn:affine}) may even not know the type of the unsafe sets, i.e., the formulation of (\ref{eqn:obstacle}). We can learn feasibility and robustness for some type-known unsafe sets with the FGO algorithm, and then use these unsafe sets to approximate any other types of unsafe sets.

\section{Implementation and Case Studies}
\label{sec:sim}

We implemented the FGO algorithm in MATLAB and performed simulations for a robot control problem. Suppose all the obstacles are of the same type but the obstacle number and their locations are unknown to the robot, and the robot is equipped with a sensor ($\frac{2}{3}\pi$ field of view (FOV) and $7m$ sensing distance with $1m$ sensing uncertainty) to detect the obstacles.

The robot dynamics are defined in the form:
\begin{equation}\label{eqn:robot}
\underbrace{\left[\begin{array}{c} 
	\dot x\\
	\dot y\\
	\dot \theta\\
	\dot v
	\end{array} \right]}_{\bm {\dot x}}=
\underbrace{\left[\begin{array}{c}  
	v\cos(\theta)\\
	v\sin(\theta)\\
	0 \\
	0
	\end{array} \right]}_{f(\bm x)}  + 
\underbrace{\left[\begin{array}{cc}  
	0 & 0\\
	0 & 0\\
	1 & 0\\
	0 & 1
	\end{array} \right]}_{g(\bm x)}\underbrace{\left[\begin{array}{c}  
	u_{1}\\
	u_{2}
	\end{array} \right]}_{\bm u}
\end{equation}
where $x, y$ denote the location along $x, y$ axis, respectively, $\theta$ denotes the heading angle of the robot, $v$ denotes the linear speed, and $u_1, u_2$ denote the two control inputs for turning and acceleration, respectively. 

We consider cost (\ref{eqn:gcost}) as the energy consumption in the form:
\begin{small}
	\begin{equation}\label{eqn:robotobj}
	J(\bm u(t)) \!=\! \int_{0}^{t_f} \eta\frac{\max\{u_{2,min}^2, u_{2,max}^2\}}{\max\{u_{1,min}^2, u_{1,max}^2\}}u_1^2(t)\! + \!(1\!-\!\eta)u_2^2(t) dt
	\end{equation}
\end{small}where $u_{1,min} < 0, u_{1,max}> 0$, $u_{2,min}< 0, u_{2,max}> 0$ denote the minimum and maximum turning control and acceleration control, respectively. $\eta \in[0,1]$ denotes a weight factor which captures the tradeoff between the two components.

We also want the robot to arrive at a destination $(x_d, y_d)\in\mathbb{R}^2$, i.e., drive $(x(t),y(t))$ to $(x_d, y_d), \forall t\in[t',t_f], t'\in[0,t_f]$, as defined in (\ref{eqn:target}). The dynamics (\ref{eqn:robot}) are not full state linearizable \cite{Khalil2002} and the relative degree of the position (output) is 2. Therefore, we cannot directly apply a CLF. However, the robot can arrive at the destination if its heading angle $\theta$ stabilizes to the desired direction and its speed $v$ stabilizes to a desired speed $v_0 > 0$, i.e.,
\begin{equation}\label{eqn:robotdst}
\begin{aligned}
\theta(t) \rightarrow \arctan(\frac{y_d - y(t)}{x_d - x(t)}), \;\;
v(t)\rightarrow v_0, \forall t\in[0,t_f].
\end{aligned}
\end{equation}
Now, we can apply the CLF method to (\ref{eqn:robotdst}) (as introduced in Def. \ref{def:clf}) since the relative degrees of the heading angle and speed are 1.

The unsafe sets (\ref{eqn:obstacle}) are defined as circular obstacles:
\begin{small}
	\begin{equation}\label{eqn:robotobs}
	\sqrt{(x(t) - x_i)^2 + (y(t) - y_i)^2} \geq r, \forall t\in[0,t_f], \forall i\in S,
	\end{equation}
\end{small}where $(x_i, y_i)$ denotes the location of the obstacle $i$, and $r > 0$ denotes the safe distance to the obstacle.

The speed and control constraints (\ref{eqn:control}) are defined as:
\begin{equation}\label{eqn:robotcontrol1}
\begin{aligned}
v_{min} \leq v(t) \leq v_{max}, \forall t\in[0,t_f],\\
u_{1,min} \leq u_1(t) \leq u_{1,max},\forall t\in[0,t_f],\\ u_{2,min} \leq u_2(t) \leq u_{2,max},\forall t\in[0,t_f].
\end{aligned}
\end{equation}
$v_{min}\geq 0, v_{max}> 0$ denote the minimum and maximum speed, respectively. The simulation parameters are listed in Table \ref{tab:para}.

\begin{table}
	\caption{Simulation Parameters}
	\label{tab:para}
	\centering
	\begin{tabular}{|c||c||c||c||c||c|}\hline
		Name     & value     & unit & Name     & value     & unit \\\hline\hline
		$p$ & 1  & unitless   &$r$     & 7       & $m$    \\\hline
		$\epsilon$     & 10 & unitless & $\Delta t$     & 0.1      & $s$   \\\hline
		$v_{min}$ & 0 &  $m/s$ & $v_{max}$ & 2 &  $m/s$  \\\hline
		$u_{1, min}$ & -0.2 &  $rad/s$ & $u_{1,max}$     & 0.2      & $rad/s$ \\\hline $u_{2, min}$ & -0.5 &  $m/s^2$ &$u_{2,max}$     & 0.5      & $m/s^2$  \\ \hline
	\end{tabular}
\end{table}

\begin{figure}[thpb]
	\centering
	\includegraphics[scale=0.45]{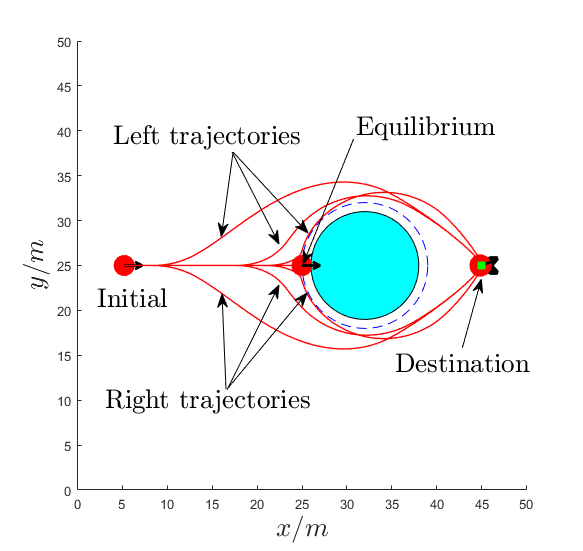}
	\caption{ FGO algorithm pre-training map with some feasible example trajectories. }
	\label{fig:train}%
\end{figure}

 We set up the FGO algorithm training environment with the initial position of the robot, the location of the obstacle and the destination as ($5m, 25m$), ($32m, 25m$) and ($45m, (25 + \varepsilon) m$) where $\varepsilon \in \mathbb{R}$, respectively. The initial heading angle and speed of the robot are 0 $deg$ and $v_{max}$, respectively, $\Delta \mathfrak{t} = 0.1, \bm\nu_{max} = -\bm\nu_{min} = (0.1,0.1,0.1,0.1)$. The map for FGO training is shown in Fig. \ref{fig:train}.

Note that the value of $\varepsilon$ in this example will affect the trajectory of the robot since we have a circular obstacle. If $\varepsilon = 0$, the robot will eventually stop at the equilibrium point shown in Fig. \ref{fig:train} since the desired heading angle (\ref{eqn:robotdst}) in the CLF exactly passes through the origin of the obstacle. If $\varepsilon > 0$, the robot goes left around the obstacle as shown in Fig. \ref{fig:train}. Otherwise, the robot turns right and then goes to the destination. 

We choose a very small $\varepsilon \ne 0$ in our FGO algorithm. Since the obstacle constraint (\ref{eqn:robotobs}) is with relative degree 2 with respect to system (\ref{eqn:robot}), we have $\bm p = (p_1, p_2), \bm q = (q_1, q_2)$. We randomly sample $M$ ($M$ is a positive integer) training and 1000 testing samples (around half of them are feasible for this robot path planning problem) for $\bm p$ and $ \bm q$ over interval $(0,3]$ and $(0,2]$, respectively.

The classification model is the support vector machine (SVM) with polynomial kernel of degree 7, i.e., the kernel function $k(\bm y,\bm z)$ is defined as
\begin{equation}\label{eqn:kernel}
k(\bm y,\bm z) = (c_1 + c_2\bm y ^T \bm z)^7.
\end{equation}
where $\bm y, \bm z$ denote input vectors of SVM (i.e., $\bm y := (\bm p, \bm q)$, as well as for $\bm z$). We set $c_1 = 0.8, c_2 = 0.5$, and the comparisons between FGO and GD are shown in Table \ref{tab:comp}.

\begin{table*}
	\caption{Comparisons between the GD and FGO algorithms}
	\label{tab:comp}
	\centering
	\begin{tabular}{|c||c||c||c||c||c||c||c||c||c|}\hline
		items     & GD     & \multicolumn{8}{c|}{FGO}\\\hline
		Training sample number $M$  &  &500 & 1000 & 1500 & 2000 & 2500 & 3000 & 3500 & 4000  \\\hline\hline
		Classification accuracy & & 0.879  & 0.927   &0.939     & 0.953     & 0.960 &0.963 &0.966 &  0.970\\\hline
		Better than GD percentage    & & 0.210 & 0.248 & 0.254     & 0.252      & 0.244  &0.282 &0.288 & 0.266 \\\hline
		Worse than GD percentage & & 0.270 &  0.190 & 0.232 & 0.204 &  0.218 & 0.218&0.240 & 0.240 \\\hline
		$\mathcal{D}_{min}/m$ (samples min.: 5.0) & 4.6 & 4.6 & 4.6   & 4.6 &  4.8 & 4.6&4.6 &4.6 & 4.6 \\\hline
	\end{tabular}
\end{table*}

\begin{figure*}[htbp]
	\centering
	
	\subfigure[FGO and GD algorithm search paths in 2D.]{
		\begin{minipage}[t]{0.48\linewidth}
			\centering
			\includegraphics[scale=0.42]{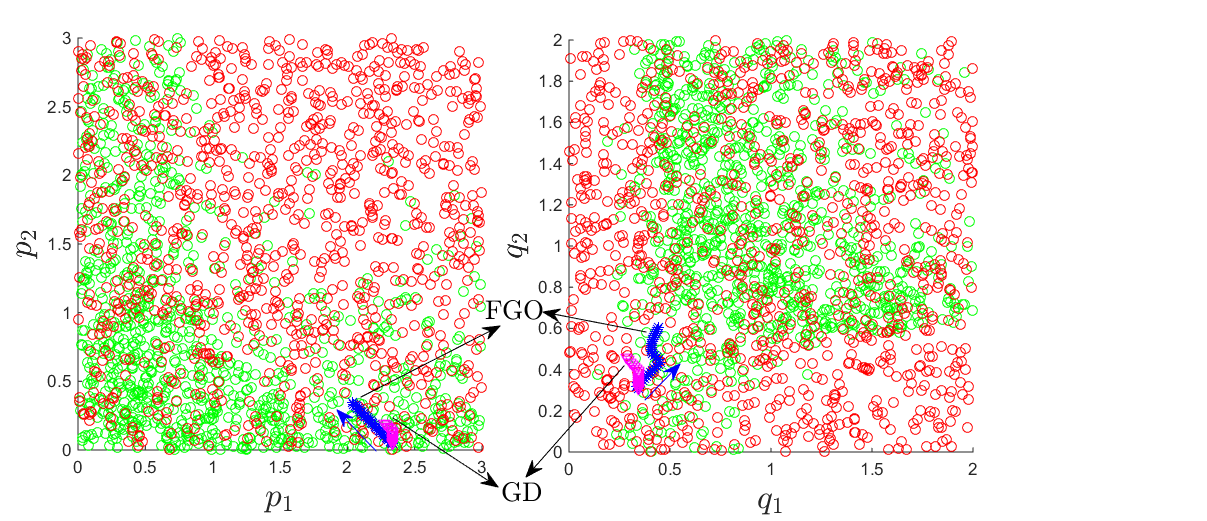}
			\label{fig:2D}
		\end{minipage}%
	}
	\subfigure[FGO and GD algorithm search paths in 3D.]{
		\begin{minipage}[t]{0.48\linewidth}
			\centering
			\includegraphics[scale=0.38]{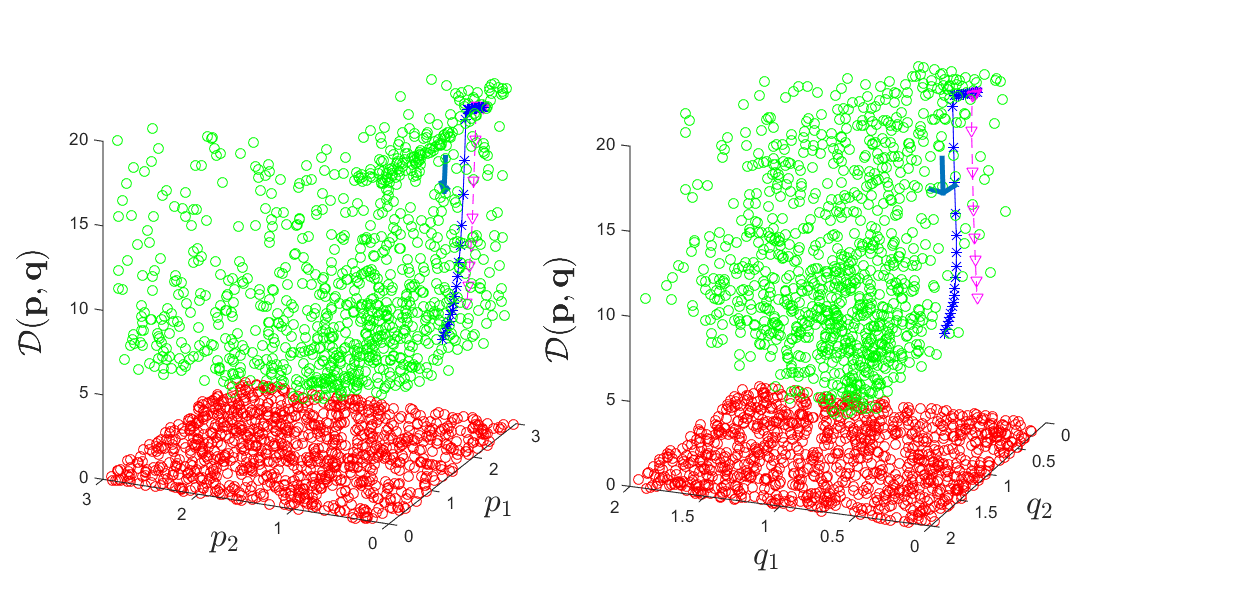}
			\label{fig:3D}
		\end{minipage}%
	}
	
	\centering
	\caption{FGO and GD algorithm search implementation. The red circles denote infeasible points and the green circles denote feasible points for $\bm p, \bm q$ in the training samples.}
\end{figure*}

The FGO has better performance compared with GD in finding $\mathcal{D}_{min}$ when the number of training samples $M$ for the hypersurface (\ref{eqn:hyperder}) is large enough, as shown in Table \ref{tab:comp}. But this advantage decreases when the classification accuracy of the hyper surface (\ref{eqn:hyperder}) further increases, which may be due to over-fitting. One comparison example between FGO and GD search paths is shown in Fig. \ref{fig:2D}, \ref{fig:3D}. We can combine the FGO and GD algorithms, i.e., we choose the best result from the FGO and GD algorithms by implementing both of them. If we continue to apply the FGO algorithm to the good results from GD, the further improvement percentage is around $5\%$ among all the testing samples. 

We have also implemented the learned optimal penalties and powers ($(p_1^*, p_2^*, q_1^*, q_2^*) = ( 0.7426, 1.9745, 1.9148$, $0.7024)$, not unique) in the definition of all the HOCBFs for all obstacles in a robot exploration problem in an unknown environment. All the circular obstacles (with different size to test the robustness of the penalty method with the learned optimal penalties and powers) move randomly. It is assumed that any obstacles that are detected by the robot will stop moving until the robot moves away. This is to ensure that the robot will not collide with obstacles passively (i.e., collisions happen due to the movement of obstacles). The robot can safely avoid all the obstacles and arrive at its destination if the obstacles do not form traps such that the robot has no way to escape.

\begin{figure}[thpb]
	\centering
	\includegraphics[scale=0.45]{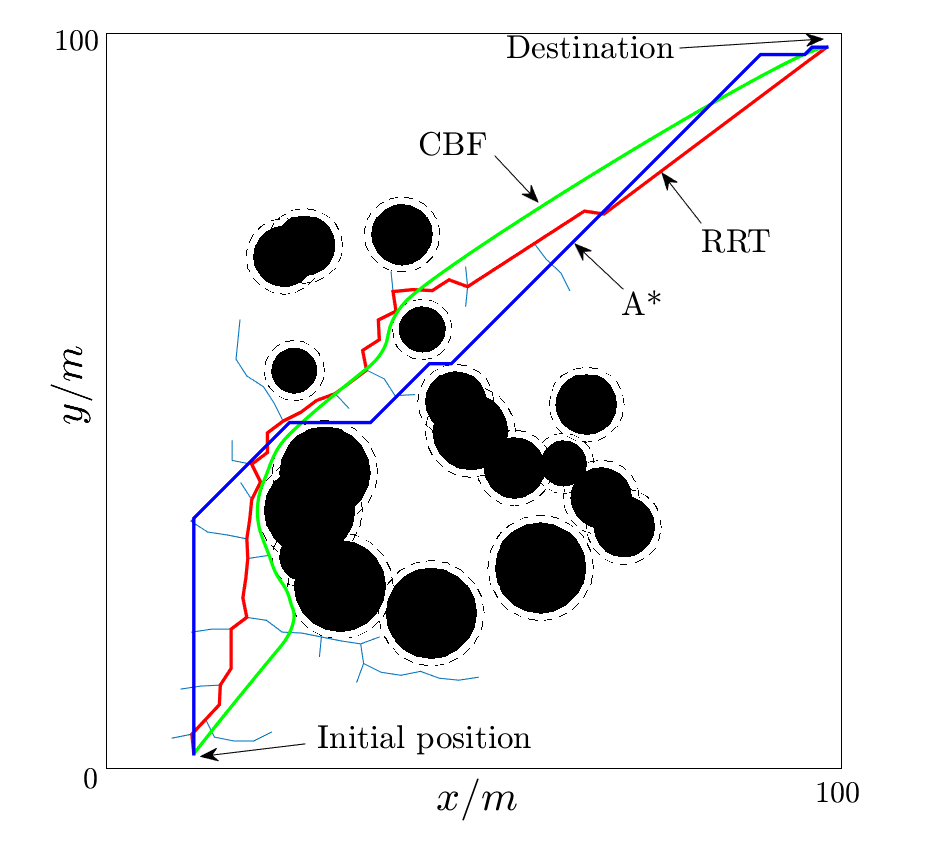}
	\caption{ Comparison of robot paths between CBF, A* and RRT. }
	\label{fig:case1}%
\end{figure}

We also compared the CBF-based robot exploration framework with the RRT and A* algorithms by picking one frame from the last simulation and fixing the location of all the obstacles, as shown in Fig. \ref{fig:case1}. Both the RRT and A* algorithms have global environment information such that they tend to choose shorter-length trajectories compared with the CBF method. But this advantage may disappear if the environment is changing fast, in which case the CBF method tends to be more robust and computationally efficient. Comparisons based on four different criteria are shown in Table \ref{tab:perf}. The computation time for the CBF method is only shown for the solution of one-step QPs in Table \ref{tab:perf} since it does not need a receding horizon, while the computation time for the RRT and A* algorithms are for the path planning time. In a dynamic environment, the RRT and A* algorithms need to re-plan their path at each time step, which may take less time than the ones shown in Table \ref{tab:perf} but is more demanding for these two algorithms. Therefore, we can see that the CBF-based framework is more adaptive in robot safe exploration.

\begin{table}
	\caption{ Performance comparison between CBF, A* and RRT in highly dynamic unknown environment}
	\label{tab:perf}
	\centering
	\begin{tabular}{|c||p{1.0cm}<{\centering}||p{1.1cm}<{\centering}||p{1.4cm}<{\centering}||p{1.4cm}<{\centering}|}\hline
		item    & R.T. compute time      & safety guarantee     & Environment knowledge  & pre-training \\\hline\hline
		CBF & $<0.01s$    &Yes    & not required     &required   \\\hline
		A*     & $1.3s$  & No     &  required  &not required     \\\hline
		RRT & $0.3s$  & No &  required & not required  \\\hline
		
	\end{tabular}
\end{table}

\section{Conclusions}
\label{sec:con}

We improved the constrained optimal control problem feasibility by maximizing the feasibility robustness through the learning of optimal parameters in the definition of a high order control barrier function that works for arbitrary relative degree constraints. This is achieved by a feasibility-guided learning approach. The proposed feasibility-guided learning approach has shown an improved ability to determine the optimal parameters compared with the gradient-descent method. The implementation on a robot safe exploration problem has shown good potential and adaptivity of the proposed framework for planning with safety guarantees compared with other path planning algorithms. Future work will focus on how to deal with traps formed by obstacles, including environment and system noise.

\bibliographystyle{plain}
\bibliography{CBF}

\begin{thebibliography}{10}

\bibitem{Aaron2012}
A.~D. Ames, K.~Galloway, and J.~W. Grizzle.
\newblock Control lyapunov functions and hybrid zero dynamics.
\newblock In {\em Proc. of 51rd IEEE Conference on Decision and Control}, pages
  6837--6842, 2012.

\bibitem{Aaron2014}
A.~D. Ames, J.~W. Grizzle, and P.~Tabuada.
\newblock Control barrier function based quadratic programs with application to
  adaptive cruise control.
\newblock In {\em Proc. of 53rd IEEE Conference on Decision and Control}, pages
  6271--6278, 2014.

\bibitem{Aubin2009}
J.~P. Aubin.
\newblock {\em Viability theory}.
\newblock Springer, 2009.

\bibitem{Boyd2004}
S.~P. Boyd and L.~Vandenberghe.
\newblock {\em Convex optimization}.
\newblock Cambridge university press, New York, 2004.

\bibitem{Carpentier2017}
J.~Carpentier, R.~Budhiraja, and N.~Mansard.
\newblock Learning feasibility constraints for multi-contact locomotion of
  legged robots.
\newblock In {\em Robotics: Science and Systems}, Cambridge, MA, 2017.

\bibitem{Freeman1996}
R.~A. Freeman and P.~V. Kokotovic.
\newblock {\em Robust Nonlinear Control Design}.
\newblock Birkhauser, 1996.

\bibitem{Galloway2013}
K.~Galloway, K.~Sreenath, A.~D. Ames, and J.W. Grizzle.
\newblock Torque saturation in bipedal robotic walking through control lyapunov
  function based quadratic programs.
\newblock {\em preprint arXiv:1302.7314}, 2013.

\bibitem{Khalil2002}
H.~K. Khalil.
\newblock {\em Nonlinear Systems}.
\newblock Prentice Hall, third edition, 2002.

\bibitem{Lindemann2018}
L.~Lindemann and D.~V. Dimarogonas.
\newblock Control barrier functions for signal temporal logic tasks.
\newblock {\em IEEE Control Systems Letters}, 3(1):96--101, 2019.

\bibitem{Mnih2016}
V.~Mnih, A.~P. Badia, M.~Mirza, A.~Graves, T.~P. Lillicrap, T.~Harley,
  D.~Silver, and K.~Kavukcuoglu.
\newblock Asynchronous methods for deep reinforcement learning.
\newblock {\em preprint arXiv:1602.01783}, 2016.

\bibitem{Mnih2015}
V.~Mnih, K.~Kavukcuoglu, D.~Silver, and A.~A.~Rusu et.al.
\newblock Human-level control through deep reinforcement learning.
\newblock {\em Nature}, 518, 2015.

\bibitem{Nguyen2016}
Q.~Nguyen and K.~Sreenath.
\newblock Exponential control barrier functions for enforcing high
  relative-degree safety-critical constraints.
\newblock In {\em Proc. of the American Control Conference}, pages 322--328,
  2016.

\bibitem{Nillson2018}
P.~Nillson and A.~D. Ames.
\newblock Barrier functions: Bridging the gap between planning from
  specifications and safety-critical control.
\newblock In {\em Proc. of 57th IEEE Conference on Decision and Control}, pages
  765--772, Miami, 2018.

\bibitem{Panagou2013}
D.~Panagou, D.~M. Stipanovic, and P.~G. Voulgaris.
\newblock Multi-objective control for multi-agent systems using lyapunov-like
  barrier functions.
\newblock In {\em Proc. of 52nd IEEE Conference on Decision and Control}, pages
  1478--1483, Florence, Italy, 2013.

\bibitem{Perrin2012}
N.~Perrin, O.~Stasse, L.~Baudouin, F.~Lamiraux, and E.~Yoshida.
\newblock Fast humanoid robot collision-free footstep planning using swept
  volume approximations.
\newblock {\em IEEE Transactions on Robotics}, 28(2):427--439, 2012.

\bibitem{Prajna2007}
S.~Prajna, A.~Jadbabaie, and G.~J. Pappas.
\newblock A framework for worst-case and stochastic safety verification using
  barrier certificates.
\newblock {\em IEEE Transactions on Automatic Control}, 52(8):1415--1428, 2007.

\bibitem{Sontag1983}
E.~Sontag.
\newblock A lyapunov-like stabilization of asymptotic controllability.
\newblock {\em SIAM Journal of Control and Optimization}, 21(3):462--471, 1983.

\bibitem{Mohit2018}
M.~Srinivasan, S.~Coogan, and M.~Egerstedt.
\newblock Feasibility envelopes for metric temporal logic specifications.
\newblock In {\em Proc. of 57th IEEE Conference on Decision and Control}, pages
  1991--1996, Miami Beach, FL, 2018.

\bibitem{Wieland2007}
P.~Wieland and F.~Allgower.
\newblock Constructive safety using control barrier functions.
\newblock In {\em Proc. of 7th IFAC Symposium on Nonlinear Control System},
  2007.

\bibitem{Wisniewski2013}
R.~Wisniewski and C.~Sloth.
\newblock Converse barrier certificate theorem.
\newblock In {\em Proc. of 52nd IEEE Conference on Decision and Control}, pages
  4713--4718, Florence, Italy, 2013.

\bibitem{Xiao2019}
W.~Xiao and C.~Belta.
\newblock Control barrier functions for systems with high relative degree.
\newblock In {\em Proc. of 58th IEEE Conference on Decision and Control}, 2019.
\newblock available in arXiv:1903.04706.

\bibitem{Wei2019}
W.~Xiao, C.~Belta, and C.~G. Cassandras.
\newblock Decentralized merging control in traffic networks: A control barrier
  function approach.
\newblock In {\em Proc. ACM/IEEE International Conference on Cyber-Physical
  Systems}, pages 270--279, Montreal, Canada, 2019.

\end{thebibliography}

\end{document}